\documentclass[aps,prl,twocolumn,showpacs,floatfix,preprintnumbers,nofootinbib,superscriptaddress]{revtex4}

\usepackage{graphicx}
\usepackage{color}
\usepackage{multirow}
\usepackage{amsmath}
\usepackage{epstopdf}
\usepackage{bbm}
   
\newcommand{\be}{\begin{equation}}
\newcommand{\ee}{\end{equation}}
\newcommand{\fett}[1]{\boldsymbol{#1}}
\newcommand{\nab}{\fett{\nabla}}
\newcommand{\dd}{{\rm d}}

\definecolor{darkgreen}{rgb}{0,0.5,0}
\definecolor{babypink}{rgb}{1, 0.1, 0.6}

\newcommand{\Section}[1]{{\em #1.}---}

\newcommand{\RR}{\mathcal{R}}

\newcommand{\pp} { \delta \hspace{-1.5pt} p}
\newcommand{\HL}{H_{\rm L}}
\newcommand{\HT}{H_{\rm T}}
\newcommand{\CLASS}{{\sc class}}

\begin{document}

\title{General relativistic corrections to $N$-body simulations \\ and the Zel'dovich approximation}

\author{Christian Fidler}
\affiliation{Catholic University of Louvain - Center for Cosmology, Particle Physics and Phenomenology (CP3)
2, Chemin du Cyclotron, B-1348 Louvain-la-Neuve, Belgium}

\author{Cornelius Rampf}
\affiliation{Institute of Cosmology and Gravitation, University of Portsmouth, Dennis Sciama Building, Burnaby Road, Portsmouth, PO1 3FX, United Kingdom}

\author{Thomas Tram}
\affiliation{Institute of Cosmology and Gravitation, University of Portsmouth, Dennis Sciama Building, Burnaby Road, Portsmouth, PO1 3FX, United Kingdom}

\author{Robert Crittenden}
\affiliation{Institute of Cosmology and Gravitation, University of Portsmouth, Dennis Sciama Building, Burnaby Road, Portsmouth, PO1 3FX, United Kingdom}

\author{Kazuya Koyama}
\affiliation{Institute of Cosmology and Gravitation, University of Portsmouth, Dennis Sciama Building, Burnaby Road, Portsmouth, PO1 3FX, United Kingdom}

\author{David Wands}
\affiliation{Institute of Cosmology and Gravitation, University of Portsmouth, Dennis Sciama Building, Burnaby Road, Portsmouth, PO1 3FX, United Kingdom}

\date{\today}

\begin{abstract}
The initial conditions for Newtonian $N$-body simulations are usually generated by applying the Zel'dovich approximation to the initial displacements of the particles using an initial power spectrum of density fluctuations generated by an Einstein--Boltzmann solver. We show that in most gauges the initial displacements generated in this way receive a first-order relativistic correction. 
We define a new gauge, the $N$-body gauge, in which this relativistic correction vanishes and show that a conventional Newtonian $N$-body simulation includes all first-order relativistic contributions (in the absence of radiation) if we identify the coordinates in Newtonian simulations with those in the relativistic $N$-body gauge. 
\end{abstract}


\maketitle

\Section{Introduction}
Cosmology has been flourishing during the last decade, especially due to the ever increasing precision of the Cosmic Microwave Background (CMB) anisotropy data \cite{Adam:2015rua, Ade:2015fwj}. But high quality data is of no use if the theoretical predictions can not be computed to the same precision. 
This has been made possible by using 
high precision codes solving the coupled system of Einstein--Boltzmann equations in perturbation theory \cite{Blas:2011rf, Lewis:1999bs, Pettinari:2013he, Huang:2013qua}. The next big data input for cosmology is Large Scale Structure (LSS) data \cite{Laureijs:2011gra,Maartens:2015mra,Ivezic:2008fe} and to successfully extract all the information in these new data sets, we will need to be able to compute the predicted LSS statistics to sufficient accuracy \cite{Schneider:2015yka}.

While the CMB can be well described by perturbation theory, the LSS is shaped by the fully non-linear nature of gravity,
where perturbation theory ceases to be a valid description on most scales of interest. The physics is still completely captured by the  
Einstein--Boltzmann equations, 
but solving its non-linear version at the required resolution is not feasible. Instead, it is common to use $N$-body simulations that solve the Newtonian equations of motion for cold dark matter (CDM) particles in full non-linearity \cite{Teyssier:2001cp,Springel:2005mi,Springel:2008cc}.
To transfer information about the matter density and velocity from the Einstein--Boltzmann solver to the $N$-body simulation at some given initial time, one usually displaces the $N$-body particles according to the Zel'dovich approximation \cite{Zeldovich:1969sb}  (or its second-order extension, 2LPT 
\cite{Scoccimarro:1997gr}). 

In this paper we discuss potential relativistic corrections to the initial displacements used in $N$-body simulations and identify a \emph{first-order} correction to the Zel'dovich approximation (ZA) which should be taken into account when setting initial displacements in a generic gauge. 
We identify a novel gauge in which relativistic corrections to both the 
ZA and the evolution equations used in $N$-body simulations vanish at first order and in the absence of radiation.

\Section{Gauges}
From the point of view of General Relativity (GR) there is no preferred coordinate system and computations can be done in any gauge. 
However, some gauges are more convenient for making the connection to Newtonian physics
\cite{Malik:2008im,Green:2011wc, Chisari:2011iq,Rampf:2012pu,Rampf:2014mga,Bertacca:2015mca}. 
For simplicity, we will consider only scalar perturbations in a spatially flat background, but the generalisation to curved space is straightforward. The most general line element in an arbitrary gauge 
is \cite{Malik:2008im} 
\begin{align}
  \dd s^2 =  a^2 \Big(& - (1+2\tilde A)\dd \tilde \eta^2  -2 \partial_i \tilde B \,\dd \tilde x^i \dd \tilde \eta  \nonumber \\
   & + \left[\delta_{ij}(1 +2 \tilde H_{\rm L}) - 2 D_{ij} \tilde H_{\rm T} \right] \dd \tilde x^i \dd \tilde x^j \Big) \,.
\end{align}
\vskip-0.2cm \noindent  Here $a$ is the cosmic scale factor, $D_{ij}\equiv \partial_i \partial_j - \delta_{ij}\nab^2/3$, and we have defined the scalar potential $\tilde A$, the scalar potential of the shift $\tilde B$, the trace of the spatial perturbation $\tilde H_{\rm L}$ and the trace-free spatial distortion $\tilde H_{\rm T}$.
One may fix the gauge by choosing explicit gauge conditions for a new set of coordinates $(\eta, x^i)$,
which read $\eta = \tilde\eta + T$ and $x^i = \tilde x^i + \partial^i L$.
For example,
a common gauge choice is the {longitudinal gauge} (sometimes called the conformal Newtonian gauge)
where the gauge freedom is used to set $T= \tilde B - \dot {\tilde H}_{\rm T}$ and $L = - \tilde H_{\rm T}$ such that $B=  H_{\rm T}=0$ in the new coordinates. 

The energy content of the Universe is defined by the components of the energy-momentum tensor ${T^\mu}_\nu$:
\begin{align}
{T^0}_0 &= -\sum_\alpha \tilde\rho_\alpha \equiv - \tilde\rho\,, \nonumber \\
{T^0}_i &= \sum_\alpha ( \tilde\rho_\alpha +  \tilde p_\alpha)\,\partial_i ( \tilde v_\alpha- \tilde B) \equiv ( \tilde\rho+ \tilde p)\,\partial_i (\tilde v- \tilde B)  \,, \nonumber \\
{T^i}_{j} &= \sum_\alpha \left( \tilde p_\alpha \,{\delta^i}_j -  \tilde p_\alpha \,{D^i}_j \tilde \Pi_\alpha \right) \equiv \tilde p \,{\delta^i}_j - \tilde  p \,{D^i}_j  \tilde \Pi\,,
\end{align}
\vskip-0.3cm \noindent 
where $\alpha$ runs over all species present in the Universe, and $\tilde \rho_\alpha$, $\tilde p_\alpha$, $\tilde v_\alpha$ and $\tilde \Pi_\alpha$ are the  
density, pressure, velocity potential and anisotropic stress of each species, respectively. Quantities without subscript will refer to totals as defined above.

We will be particularly interested in the class of comoving-orthogonal gauges (hereafter referred to as comoving gauges) defined by setting the shift equal to the peculiar velocity potential, i.e., $B = v$.
This uniquely fixes the temporal gauge 
with $T= \tilde B-\tilde v$, while the spatial gauge $L$ can be chosen freely.
%
The density contrast, $\delta=(\rho-\bar\rho)/\bar\rho$, is independent of the spatial gauge transformation $L$, so it is identical in all comoving gauges.
The same is true for the lapse perturbation $\xi\equiv A$.
Velocities, however,  depend on the time derivative of the spatial gauge generator, i.e., on $\dot L$. 

To first order in perturbation theory, the $(00)$, $(0i)$ and $(i \neq j)$ components of the Einstein equations are 
\begin{align}
 \!\!\!\!\nab^2 \left[ \HL+ \frac{\nab^2}{3} \HT -\frac{\dot{a}}{a} ( v - \dot{H}_{\rm T})  \right]  &= - 4\pi G \bar \rho a^2 \delta , \label{eq:Poissonfull}
 \\ 
\!\!\!\!\frac{\dot{a}}{a} \xi-\dot{H}_{\rm L} - \frac{\nab^2}{3}\dot{H}_{\rm T} &= 0 \,,  \label{eq:Q} \\
\!\!\!\xi \! +\! \HL + \!\frac{\nab^2}{3} \HT\! - \!\bigg[\frac{\partial}{\partial \eta} +2 \frac{\dot{a}}{a}\bigg] \!\big(v-\dot{H}_{\rm T} \big) \! &= 8 \pi G a^2 p \Pi \, .  \label{eq:velocity} 
\end{align}
\vskip-0.2cm \noindent The Einstein equations are supplemented by the continuity and the momentum conservation equation: 
\begin{align} \label{eq:continuity}
\bigg[\frac{\partial}{\partial \eta} + 3 \frac{\dot{a}}{a} \bigg] \bar \rho\, \delta + 3 \frac{\dot{a}}{a}\pp &= - (\rho + p)\left( \nab^2 v + 3 \dot{H}_{\rm L} \right)\,, \\
\label{eq:Navier}
(\rho + p) \,\xi &= \frac{2}{3} p \nab^2 \Pi - \pp \,,
\end{align}
\vskip-0.2cm \noindent  with the pressure perturbation $\pp$. Note that the continuity equation (\ref{eq:continuity}) holds in the same form for each individual component in a multicomponent Universe while the momentum conservation reads:
\begin{align}
\left[\frac{\partial}{\partial\eta} +4 \frac{\dot{a}}{a}\right]&(\rho_\alpha+p_\alpha)(v_\alpha-v)=  \nonumber \\ \label{eq:Naviercomp}
  &\frac{2}{3}p_\alpha\nab^2 \Pi_\alpha -\pp_\alpha - (\rho_\alpha+p_\alpha)\xi \,.
\end{align}

\vskip-0.2cm  The gauge invariant Bardeen potential $\Phi$ in comoving gauges is given by \cite{Malik:2008im}
\be
\Phi = \HL + \frac{\nab^2}{3} \HT - \frac{\dot{a}}{a} \left(v-\dot{H}_{\rm T} \right) \,.
\ee
\vskip-0.2cm \noindent Equation~(\ref{eq:Poissonfull}) can then be identified as the relativistic Poisson equation
\be
 \label{Poisson}
 \nab^2 \Phi =  - 4\pi G \bar \rho a^2 \delta \,.
\ee
\vskip-0.2cm \noindent This is identical to the Newtonian Poisson equation solved in an $N$-body simulation.
Using the continuity equation~(\ref{eq:continuity}),
 the momentum conservation~(\ref{eq:Naviercomp}) and Eq.\,(\ref{eq:velocity}),
 we find the following evolution equations for a pressureless fluid component  ($p_\alpha = \Pi_\alpha=0$, e.g., for dark matter):
\begin{align} 
 \dot{\delta}_\alpha + \nab \cdot \fett{v}_\alpha &=  - 3\dot{H}_{\rm L} \,, \label{conti} \\
\left(\frac{\partial}{\partial \eta} + \frac{\dot{a}}{a}\right) \fett{v}_\alpha &= \nab\Phi +  \nab\gamma \,, \label{eq:geo}
\end{align}
where $\fett{v}_\alpha = \nab v_\alpha$, and we have defined
\be \label{gamma}
  \gamma \equiv   \ddot H_{\rm T} + \frac{\dot{a}}{a}\dot{H}_{\rm T} 
  - 8 \pi G a^2 p \Pi  \,.
\ee
Equation~(\ref{conti}) is identical with the Newtonian continuity
equation when $\dot H_{\rm L}=0$.
The geodesic equation~(\ref{eq:geo}) agrees with the Newtonian Euler equation used to update the particle velocities in an $N$-body simulation when $\gamma$ vanishes.

The geodesic equation~(\ref{eq:geo}) requires us to know the potential, $\Phi$, and we have seen that this can be obtained from the Poisson equation~(\ref{Poisson}) if we can compute the comoving density. 
In a Newtonian simulation the density is computed by counting the number
of particles in a  
volume element:
\begin{equation}
\label{eq:rhocount}
\rho_\text{count} = \frac{1}{a^3} \sum \limits_{\text{particles}} m \,\delta_{\rm D}^{(3)}(\fett{x}-\fett{x}_p)\,.
\end{equation}
\vskip-0.2cm \noindent 
By contrast, the relativistic density, $\rho$, has to take into account the inhomogeneous deformation of space.
The trace of the 3-metric, $\HL$, modifies the volume by a factor of $(1+ 3\HL)$, while $\HT$ leaves the volume unchanged: 
\begin{equation} \label{rhosim}
  \rho =  (1-3 \HL) \rho_{\text{count}} \,.
\end{equation}
\vskip-0.2cm \noindent  This means that even though the Poisson equation is formally identical to its Newtonian counterpart, the density in the simulation is not necessarily the comoving density required by the relativistic Poisson equation.

Let us define the gauge in which the counting density matches the comoving density 
by requiring a vanishing $\HL$. 
This fixes the spatial gauge transformation: 
\vskip-0.4cm 
\be
 \nab^2 L = 3 \tilde H_{\rm L}-3 \frac{\dot{a}}{a} ( \tilde B - \tilde v )\,.
\ee
\vskip-0.2cm \noindent 
In the following we shall call this the \emph{N-body gauge}.
In this gauge,
the continuity equation~(\ref{conti}) has the Newtonian form and
 the Poisson equation solved in an $N$-body simulation 
is consistent with GR, since the computed density matches the
comoving density to first order. 
However there is a potential correction to the geodesic equation~(\ref{eq:geo}) from $\gamma$. We will now demonstrate that this correction vanishes in matter/$\Lambda$ domination. 

Equation~(\ref{eq:Navier}) relates the lapse perturbation, $\xi$, directly to the anisotropic stress and pressure perturbation. 
This implies that $\xi$ vanishes in any comoving gauge when $\pp = \Pi =0$.
Equation~(\ref{eq:Q}) then requires that $\dot{\RR}=0$
 where we identify the comoving curvature perturbation
\be
\label{def:R}
\RR \equiv \HL + \frac{\nab^2}{3} \HT \,.
\ee
\vskip-0.2cm \noindent 
In the $N$-body gauge ($\HL=0$), this implies that $\HT$ is constant and therefore $\gamma$ vanishes when $\delta p = \Pi =0$.

Another popular comoving gauge choice is the total matter (TOM) gauge \cite{Malik:2008im}
in which the metric potential $\HT$ is set to zero but $\HL\neq0$. 
In the absence of anisotropic stress ($\Pi=0$) there are no corrections to the classical Euler equation, while the Poisson equation is unmodified in all the comoving gauges. 
However, the counting density in an $N$-body simulation (\ref{eq:rhocount}) would not match the comoving density due to the volume deformation if $\HL\neq0$, leading to relativistic corrections. 

We conclude that the $N$-body gauge is uniquely suited for $N$-body simulations, with GR corrections appearing at most at second order in the evolution equations. 
Thus, although it has not previously been noted in the literature, conventional Newtonian $N$-body simulations actually use initial displacements corresponding to those in the $N$-body gauge.

\Section{The Zel'dovich approximation}
The ZA is the first-order solution for the Lagrangian displacement field, $\fett{\psi}$. We use the Lagrangian map $\fett{q} \mapsto \fett{x}(\fett{q},\eta)$, where
\be 
 \fett{x}(\fett{q},\eta) = \fett{q} + \fett{\psi}(\fett{q},\eta)
\ee
 denotes the trajectory of a fluid particle from its initial position $\fett{q}$ to its subsequent coordinate position at time $\eta$.
The velocity is the (Lagrangian) time derivative of the position,
$\fett{v}(\fett{x}(\fett{q},\eta)) =  \dot{\fett{x}}(\fett{q},\eta)$, or simply
\begin{equation}\label{eq:DispDef}
 \fett{v}_\alpha = \dot{\fett{\psi}}_\alpha \,,
\end{equation}
\vskip-0.2cm \noindent 
where the (peculiar) velocity $\fett{v}_\alpha$ obeys the geodesic equation for pressureless matter.
The continuity equation for the matter over-density $\delta_\alpha$ in Newtonian theory reads
\be \label{eq:BoltzMatter}
\dot{\delta}_\alpha + \nab \cdot \fett{v}_\alpha = 0\,.
\ee
\vskip-0.2cm \noindent 
In the infinite past 
the displacement is zero so that the distribution of matter is uniform,
hence integrating Eqs.\,(\ref{eq:DispDef})--(\ref{eq:BoltzMatter}) we find the well-known ZA
\begin{equation}
-\nab \cdot \fett{\psi}_\alpha = \delta_\alpha \,.  \label{eq:ZelMD}
\end{equation}

\vskip-0.2cm  The derivation of the ZA assumes a Newtonian continuity equation, but for a general gauge choice the corresponding continuity equation~(\ref{conti}) includes a relativistic correction. Typically (e.g., in TOM or longitudinal gauge) these corrections vanish during matter domination, but the ZA is computed by integrating over the whole past history of the Universe, including the preceding period of radiation domination.

Therefore we derive the relativistic ZA using the continuity equation~(\ref{conti}) for the matter components. In GR the density changes due to two different effects. First particle movement generates over- and under-dense regions which is captured by the velocity divergence, $\nab \cdot \fett{v}_\alpha$, as in Newtonian theory. But in addition space can be deformed in GR in an inhomogeneous way, described by the metric term $\dot{H}_{\rm L}$ on the right-hand side of Eq.\,(\ref{conti}), which creates over- or under-dense regions without requiring any particle movement. In contrast to the density, the displacement field only traces the movement of particles.
Combining Eqs.\,(\ref{eq:DispDef}) and the relativistic continuity equation~(\ref{conti}), and 
integrating over the past history of the Universe starting from a homogeneous distribution we obtain the ZA including the GR correction:
\begin{align} \label{eq:modDispl}
-\nab \cdot \fett{F}_\alpha &= \delta_\alpha + 3 \HL \,,
\end{align}
\vskip-0.2cm \noindent  where $\fett{F}_\alpha$ is the relativistic displacement field.
The counting density is then given by $\rho_{\text{count}, \alpha}\!=\!(\bar \rho_{0, \alpha}/a^3)(1-\nab \cdot \fett{F}_\alpha)$. Using~(\ref{eq:modDispl}) we then obtain:
\begin{equation} \label{rhosim2}
\rho_{\text{count}, \alpha} = \frac{\bar \rho_{0, \alpha}}{a^3}(1+\delta_\alpha + 3\HL)\,.
\end{equation}
\vskip-0.2cm \noindent  Substituting this last expression into Eq.\,\eqref{rhosim} we recover
\begin{equation} \label{rhotom}
\rho_\alpha = \frac{\bar{\rho}_{0, \alpha}}{a^3}(1+\delta_\alpha)\,,
\end{equation}
which is the definition of the comoving density contrast. 

We have thus shown that if we generate initial conditions for $N$-body simulations with the relativistic displacement~(\ref{eq:modDispl}), the distribution of particles correctly reproduces the comoving gauge density.
As a consequence, this relativistic correction should be included in the initial displacement in an arbitrary gauge.
\begin{figure}
\includegraphics[width=\columnwidth]{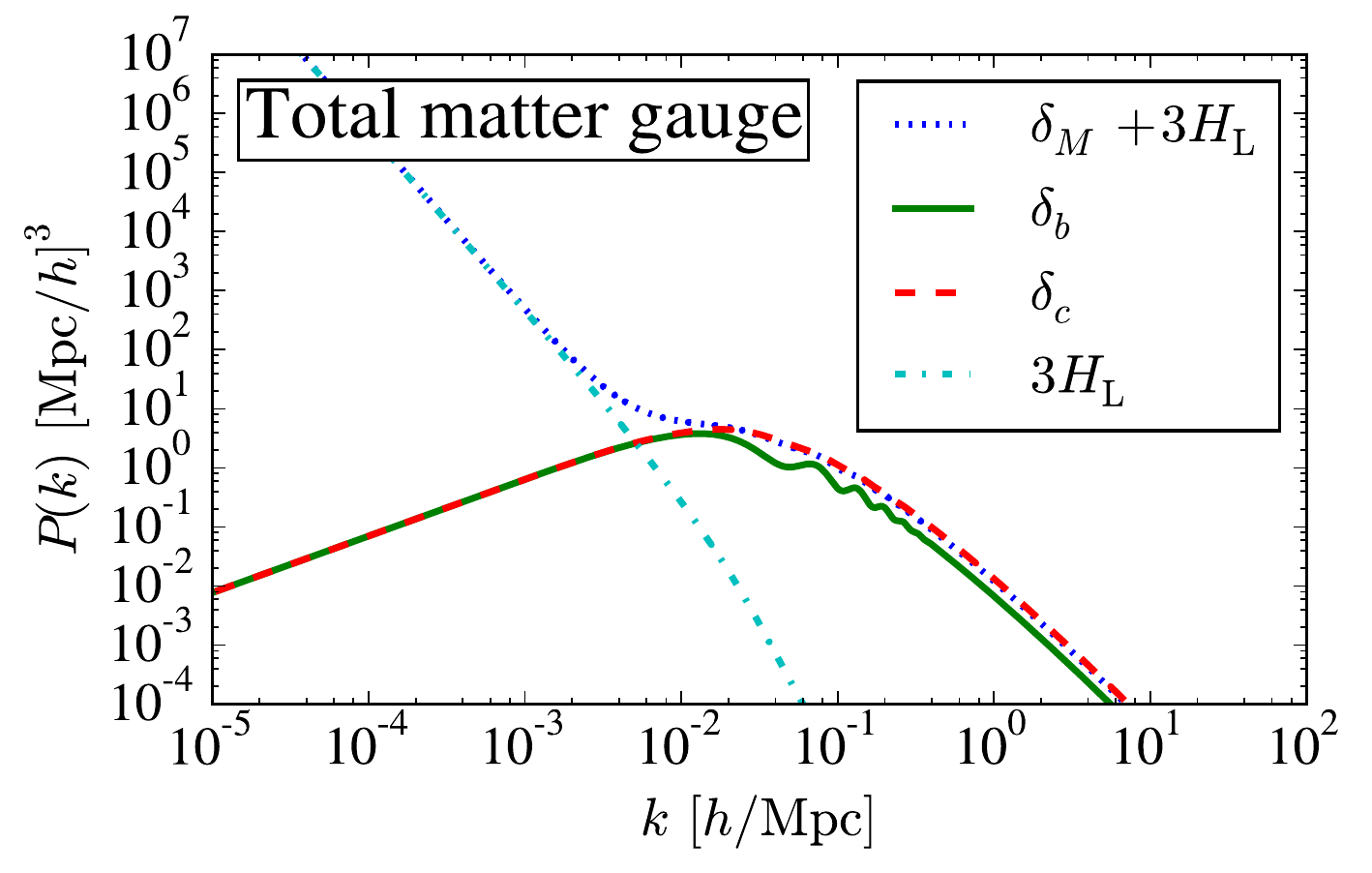}
\caption{\label{fig:potentials}To illustrate the GR correction to the initial conditions, we plot the power spectrum of $\nab \cdot \fett{F}_\alpha$  (blue [dotted] line) according to equation~\eqref{eq:modDispl} at redshift $z=100$ in the TOM gauge. $\delta_M=\delta_c+\delta_b$ includes CDM plus baryons. We also plot the individual power spectra for CDM and baryons, as well as the power spectrum of the correction term $3 \HL$ alone. The displacement fields in TOM gauge and longitudinal gauge coincide at first order.}
\end{figure}

\begin{figure}
\includegraphics[width=\columnwidth]{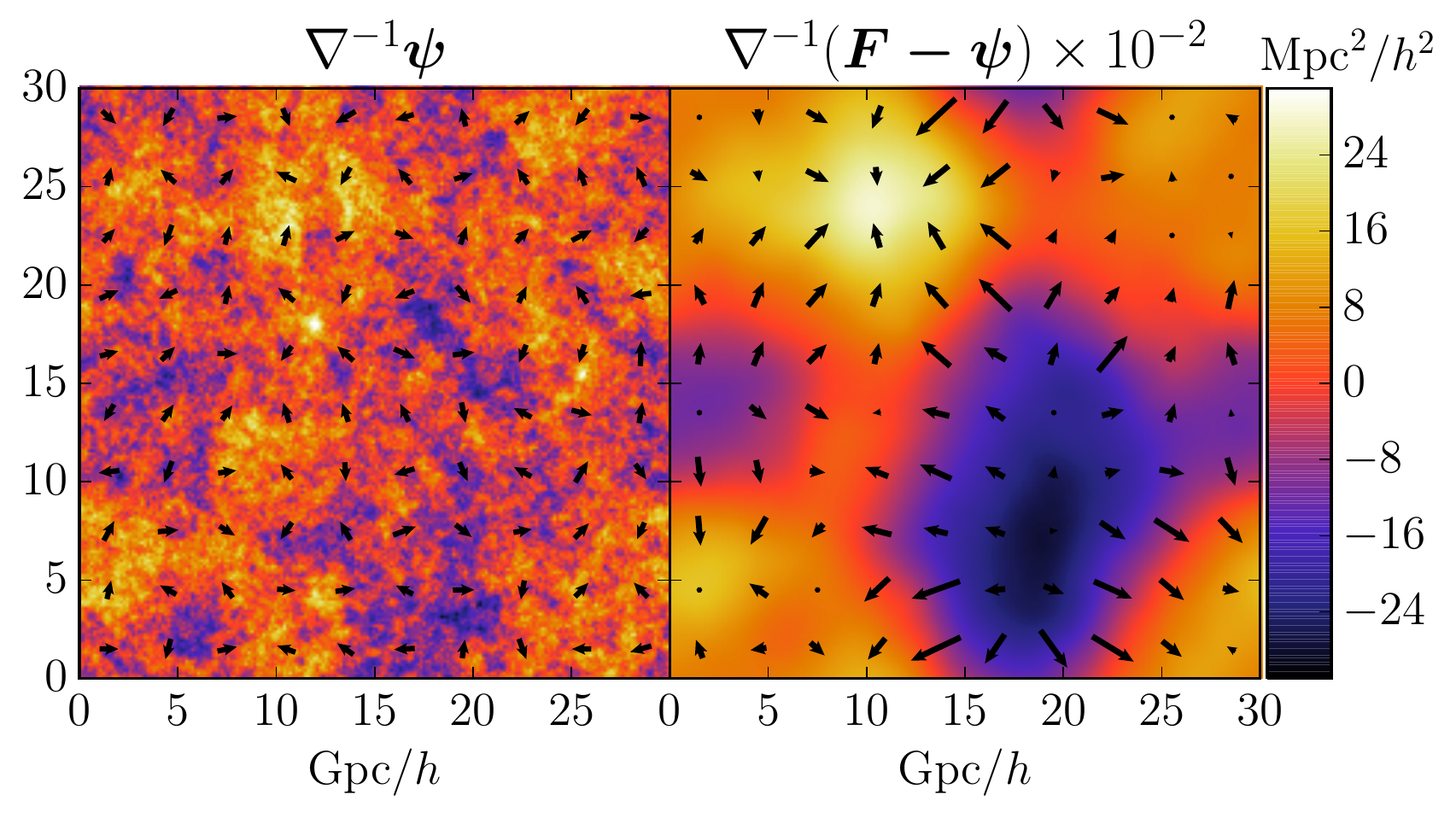}
\caption{\label{fig:displacements} We show the potential of the displacement field in TOM gauge at $z = 100$. 
The initial scalar displacements, shown as arrows, are the gradient of this field. All arrows have been multiplied by a factor $4000$ for improved visibility.
Note that since $\HT=0$ in the TOM gauge, the spatial displacements are the same as those in the longitudinal gauge at first order. The left side shows the potential for the classical Zel'dovich displacement $\fett{\psi}$, while the right side shows the relativistic correction to $\fett{\psi}$.}
\end{figure}

The impact of the correction for the CDM and baryons in TOM gauge is illustrated in Fig.~\ref{fig:potentials}. 
It shows the power spectrum of the comoving density 
compared to the power spectrum of $\HL$,
which is equal to the comoving curvature perturbation ${\cal R}$ in the TOM {gauge~($\HT\!=\!0$).} On very large scales $\HL$ dominates leading to a considerably modified initial displacement, while small scales are not affected by the relativistic correction. Fig.~\ref{fig:displacements} shows the scalar potential of the displacement field for the classical ZA in the left panel and the relativistic correction from $3\HL$ in the right panel. The displacement caused by the relativistic correction is two orders of magnitude larger than the classical ZA, but it is only present on large scales. 
In the $N$-body gauge, however, $\HL=0$ and there is no relativistic correction to the displacement.

\begin{figure}
\includegraphics[width=\columnwidth]{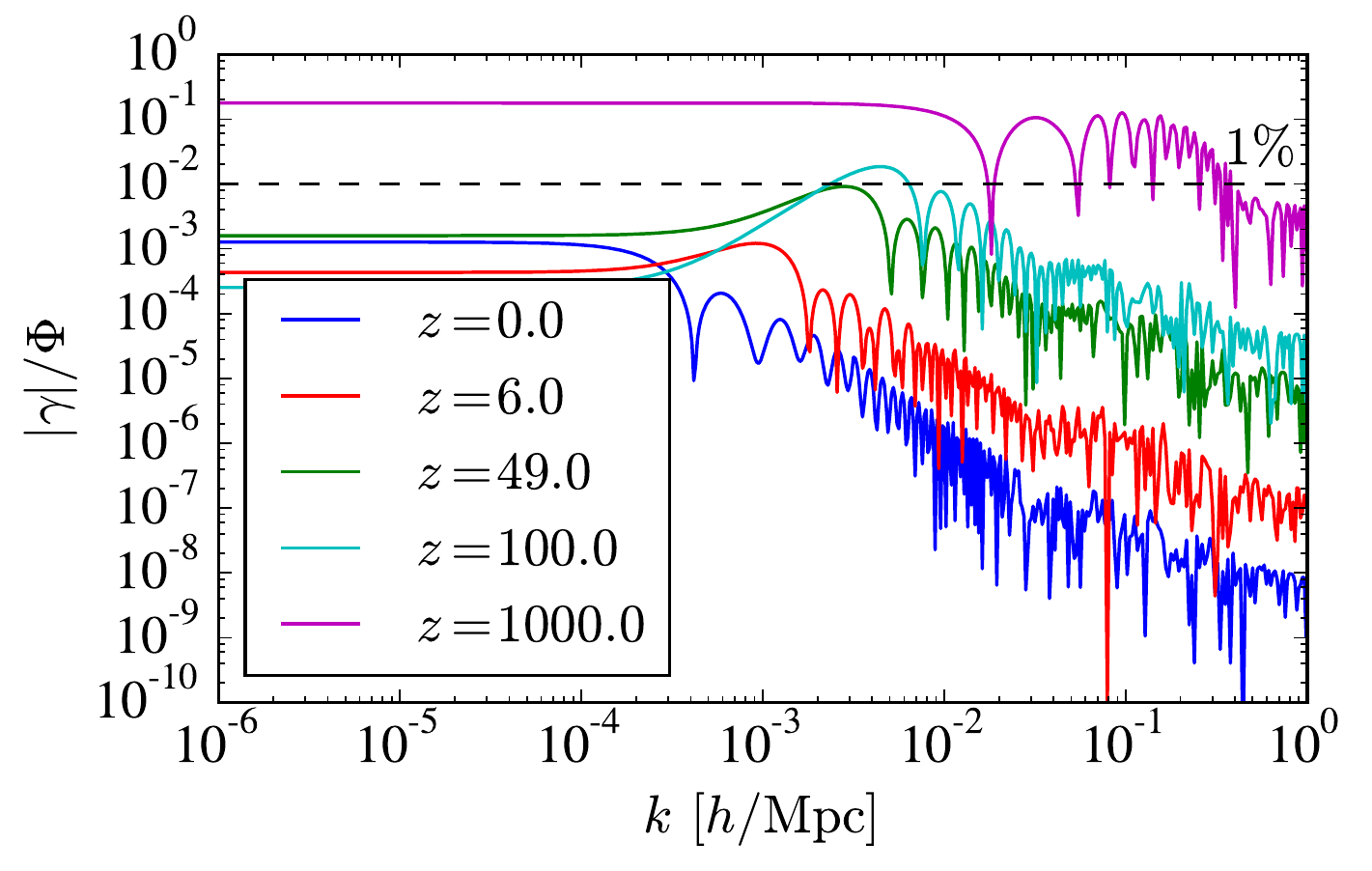}
\caption{\label{fig:xiPhiratio} 
Ratio of $|\gamma|$ in the $N$-body gauge compared to the Bardeen potential $\Phi$, illustrating the impact of radiation contaminants on conventional $N$-body simulations. 
On small scales the relevance of residual radiation is continuously decreasing in time, but on large scales, around $z \simeq 50$, there is a cancelation between competing contributions in $\gamma$.
}
\end{figure}

In a realistic cosmology there is residual radiation at high redshifts, which should be taken into account when setting up the initial conditions for $N$-body simulations in any gauge. In our $N$-body gauge only the relativistic geodesic equation is modified by the presence of radiation, described by $\gamma$ in Eq.~(\ref{eq:geo}), which is missing in conventional $N$-body simulations.
Thus to get a smooth transition from a relativistic to a Newtonian description, $N$-body simulations should not be initialised at high redshifts, when radiation is still important. 
In Fig.~\ref{fig:xiPhiratio} we show the ratio of $|\gamma|/\Phi$, describing the correction to the geodesic equation. $N$-body simulations which are initialised at redshifts higher than 49 receive larger than percent level corrections to the Euler equation initially. 
There is an inevitable tension between the need to minimise radiation corrections (that require the $N$-body start time to be at lower redshifts) with the need to reduce non-linear corrections to the initial conditions (which are minimised at early times) 
\cite{Schneider:2015yka}. The usual solution is using Newtonian 2LPT to set initial conditions at lower redshifts. However to do this while consistently including relativistic corrections has not yet been done and remains a challenge for future work.

\Section{Conclusions}
We have shown that the initial displacements for particles in an arbitrary gauge receive relativistic corrections. 
These corrections however vanish in our novel {\em N-body gauge}, where the Newtonian ZA is recovered, and the relativistic evolution equations take the Newtonian form for vanishing pressure perturbations and anisotropic stress.  
 Therefore, the initial displacements and the output of Newtonian $N$-body simulations should be understood in terms of our $N$-body gauge.
By contrast, the density that would be computed in $N$-body simulations using the TOM gauge would not agree with the comoving density, because of the relativistic volume deformation, which is absent in Newtonian simulations.

When comparing simulations to {LSS} surveys {(e.g., SKA and Euclid \cite{Laureijs:2011gra,Maartens:2015mra})}, the particle positions in the $N$-body simulation must be converted to observable coordinates \cite{observation}. This conversion depends on the gauge used and, as argued above, the $N$-body positions should be interpreted in the $N$-body gauge. 
However, some quantities do not depend on the spatial gauge used. For example, the density is identical in all comoving gauges and therefore quantities derived from it, such as the matter power spectrum, are the same in all comoving gauges.

In the commonly used longitudinal gauge, the authors of \cite{Chisari:2011iq,Rampf:2012pu} showed that there are a number of GR terms in the relativistic equations which are apparently missing in the Newtonian equations; these extra terms then rather mysteriously cancel. We have shown that in fact these additional GR terms are nothing other than the gauge transformations from quantities defined in the longitudinal to those defined in the $N$-body gauge, in which gauge the relativistic equations coincide precisely with the Newtonian ones.

Finally, let us  briefly discuss conventions for the matter power spectra in available Boltzmann codes, namely in {\sc camb}~\cite{Lewis:1999bs} and \CLASS{} \cite{Blas:2011rf}.
The matter power spectrum computed by {\sc camb} is in the synchronous gauge; this differs slightly from the $N$-body gauge matter power spectrum on large scales due to the (small) total velocity contribution from baryons, neutrinos and photons. In \CLASS{} the matter power spectrum is computed in a gauge comoving with the non-relativistic species. Again, this differs from the matter power spectrum in the $N$-body gauge on large scales.
 We have modified \CLASS{} to also output the comoving density power spectra in the $N$-body gauge for cold dark matter, baryons, the sum of cold dark matter and baryons, warm dark matter and massive neutrinos.
It is available at \url{https://Github.com/ThomasTram/NbodyCLASS}.
These densities can be used to generate the displacement field using the ZA (to first order) in the $N$-body gauge. 
A Newtonian $N$-body simulation starting from these initial conditions (or its 2LPT extension to second order) computes the relativistic evolution up to first order.

\acknowledgments

We thank Marco Bruni for useful discussions.
CF~is supported by the Wallonia-Brussels
Federation grant ARC\,11/15-040 and the Belgian Federal Office for
Science, Technical\,\&\,Cultural Affairs through the 
Interuniversity Attraction Pole\,P7/37. CR~acknowledges the support of the individual fellowship RA\,2523/1-1 
from the Deutsche Forschungsgemeinschaft.\,TT,\,RC,\,KK, and DW are supported by the UK Science \& Technology Facilities Council grants ST/K00090X/1 \& ST/L005573/1.



\begin{thebibliography}{}

\bibitem{Adam:2015rua}
  R.~Adam {\it et al.}  [Planck Collaboration],
  arXiv:1502.01582 [astro-ph.CO].

\bibitem{Ade:2015fwj} 
  P.~A.~R.~Ade {\it et al.} [BICEP2 and Keck Array Collaborations],
  Astrophys.\ J.\  {\bf 811}, no. 2, 126 (2015)
  [arXiv:1502.00643 [astro-ph.CO]].

\bibitem{Lewis:1999bs}
  A.~Lewis, A.~Challinor and A.~Lasenby,
  Astrophys.\ J.\  {\bf 538}, 473 (2000)
  [astro-ph/9911177].

\bibitem{Blas:2011rf}
  D.~Blas, J.~Lesgourgues and T.~Tram,
  JCAP {\bf 1107}, 034 (2011)
  [arXiv:1104.2933 [astro-ph.CO]].

\bibitem{Pettinari:2013he}
  G.~W.~Pettinari, C.~Fidler, R.~Crittenden, K.~Koyama and D.~Wands,
  JCAP {\bf 1304}, 003 (2013)
  [arXiv:1302.0832 [astro-ph.CO]].

\bibitem{Huang:2013qua}
  Z.~Huang and F.~Vernizzi,
  Phys.\ Rev.\ D {\bf 89}, no. 2, 021302 (2014)
  [arXiv:1311.6105 [astro-ph.CO]].

\bibitem{Laureijs:2011gra}
  R.~Laureijs {\it et al.}  [EUCLID Collaboration],
  arXiv:1110.3193 [astro-ph.CO].

\bibitem{Maartens:2015mra} 
  R.~Maartens {\it et al.} [SKA Cosmology SWG Collaboration],
  PoS AASKA {\bf 14}, 016 (2015)
  [arXiv:1501.04076 [astro-ph.CO]].

\bibitem{Ivezic:2008fe}
  Z.~Ivezic {\it et al.}  [LSST Collaboration],
  arXiv:0805.2366 [astro-ph].

\bibitem{Schneider:2015yka} 
  A.~Schneider, R.~Teyssier, D.~Potter, J.~Stadel, J.~Onions, D.~S.~Reed, R.~E.~Smith and V.~Springel {\it et al.},
  arXiv:1503.05920 [astro-ph.CO].
  
\bibitem{Teyssier:2001cp}
  R.~Teyssier,
  Astron.\ Astrophys.\  {\bf 385}, 337 (2002)
  [astro-ph/0111367].

\bibitem{Springel:2005mi}
  V.~Springel,
  Mon.\ Not.\ Roy.\ Astron.\ Soc.\  {\bf 364}, 1105 (2005)
  [astro-ph/0505010].

\bibitem{Springel:2008cc}
  V.~Springel, J.~Wang, M.~Vogelsberger, A.~Ludlow, A.~Jenkins, A.~Helmi, J.~F.~Navarro and C.~S.~Frenk {\it et al.},
  Mon.\ Not.\ Roy.\ Astron.\ Soc.\  {\bf 391}, 1685 (2008)
  [arXiv:0809.0898 [astro-ph]].

\bibitem{Zeldovich:1969sb}
  Ya.~B.~Zeldovich,
  Astron.\ Astrophys.\  {\bf 5}, 84 (1970).

\bibitem{Scoccimarro:1997gr}
  R.~Scoccimarro,
  Mon.\ Not.\ Roy.\ Astron.\ Soc.\  {\bf 299}, 1097 (1998)
  [astro-ph/9711187].

\bibitem{Malik:2008im} 
  K.~A.~Malik and D.~Wands,
  Phys.\ Rept.\  {\bf 475}, 1 (2009)
  [arXiv:0809.4944 [astro-ph]].

\bibitem{Green:2011wc} 
  S.~R.~Green and R.~M.~Wald,
  Phys.\ Rev.\ D {\bf 85}, 063512 (2012)
  [arXiv:1111.2997 [gr-qc]].

\bibitem{Chisari:2011iq}
  N.~E.~Chisari and M.~Zaldarriaga,
  Phys.\ Rev.\ D {\bf 83}, 123505 (2011)
  [Erratum-ibid.\ D {\bf 84}, 089901 (2011)]
  [arXiv:1101.3555 [astro-ph.CO]].


\bibitem{Rampf:2012pu} 
  C.~Rampf and G.~Rigopoulos,
  Mon.\ Not.\ Roy.\ Astron.\ Soc.\ Lett.\  {\bf 430}, L54 (2013)
  [arXiv:1210.5446 [astro-ph.CO]].


\bibitem{Rampf:2014mga}
  C.~Rampf and A.~Wiegand,
  Phys.\ Rev.\ D {\bf 90}, no. 12, 123503 (2014)
  [arXiv:1409.2688 [gr-qc]].

\bibitem{Bertacca:2015mca} 
  D.~Bertacca, N.~Bartolo, M.~Bruni, K.~Koyama, R.~Maartens, S.~Matarrese, M.~Sasaki and D.~Wands,
  Class.\ Quant.\ Grav.\  {\bf 32}, no. 17, 175019 (2015)
  [arXiv:1501.03163 [astro-ph.CO]].

\bibitem{observation} 
  J.~Yoo,
  Phys.\ Rev.\ D {\bf 82}, 083508 (2010)
  [arXiv:1009.3021 [astro-ph.CO]];
  C.~Bonvin and R.~Durrer,
  Phys.\ Rev.\ D {\bf 84}, 063505 (2011)
  [arXiv:1105.5280 [astro-ph.CO]];
  A.~Challinor and A.~Lewis,
  Phys.\ Rev.\ D {\bf 84}, 043516 (2011)
  [arXiv:1105.5292 [astro-ph.CO]];
  D.~Jeong and F.~Schmidt,
  Class.\ Quant.\ Grav.\  {\bf 32}, no. 4, 044001 (2015)
  [arXiv:1407.7979 [astro-ph.CO]].


\end{thebibliography}
\end{document}